\documentclass[prl,preprintnumbers,tightenlines,twocolumn,superscriptaddress]{revtex4-1}
\usepackage{braket}
\usepackage{bbold}
\usepackage{dsfont}
\usepackage{amsmath}
\usepackage{amssymb}
\usepackage{amsfonts}
\usepackage{mathtools}
\usepackage{graphicx}% Include figure files
\usepackage{dcolumn}% Align table columns on decimal point
\usepackage{bm}% bold math
\usepackage{multirow}
\usepackage{leftidx}
\usepackage{color}
\usepackage{MnSymbol}
\usepackage[mathscr]{eucal}
\usepackage[german,english]{babel}
\usepackage{array}
\usepackage{pifont}%
\usepackage[capitalize]{cleveref}
\usepackage[utf8]{inputenc}

\begin{document}
\title{Probing ultracold gases using photoionization fine structure}
\author{P. Giannakeas}
\email{pgiannak@pks.mpg.de}
\affiliation{Max-Planck-Institut f\"ur Physik komplexer Systeme, N\"othnitzer Str.\ 38, D-01187 Dresden, Germany }
	
\author{Matthew T. Eiles}
% \email{meiles@pks.mpg.de}
\affiliation{Max-Planck-Institut f\"ur Physik komplexer Systeme, N\"othnitzer Str.\ 38, D-01187 Dresden, Germany }
\author{L. Alonso}
% \email{robichf@purdue.edu}
\affiliation{Max Planck Institute for Biogeochemistry, Hans-Knöll-Str.10, D-07745 Jena, Germany }

\author{F. Robicheaux}
% \email{robichf@purdue.edu}
\affiliation{Department of Physics and Astronomy, Purdue University, 47907 West Lafayette, IN, USA}
\affiliation{ Purdue Quantum Science and Engineering Institute, Purdue University, West Lafayette, Indiana 47907, USA}
\author{Jan M. Rost}
% \email{rost@pks.mpg.de}
\affiliation{Max-Planck-Institut f\"ur Physik komplexer Systeme, N\"othnitzer Str.\ 38, D-01187 Dresden, Germany }

\begin{abstract}
Photoionization of atoms immersed in an environment such as an ultracold gas is investigated.
We show that the interference of two ionization pathways, one passing directly to the continuum and one accounting for scattering processes between the photoelectron and a neighboring atom, produces a fine structure in the photoionization cross-section over an energy range less than 1 eV above threshold.
This fine structure includes all the details of the corresponding three-body system, e.g. the interatomic distance or the scattering information of the electron-atom subsystem; therefore, photoelectrons produced in a multi-particle environment can be utilized as structural probes.
As an illustration, for experimentally relevant parameters, we propose a scheme based on the photoionization of a Rydberg molecule where the low-energy electron-atom phase shifts are extracted from the fine structure spectra using neural networks.
\end{abstract}
\maketitle

Spectroscopic imaging tools infer many-body effects by measuring the interaction-induced shift of transition lines away from their known single particle values.  
Such a technique was used to observe the first hydrogen Bose-Einstein condensate (BEC) by measuring the shift, proportional to density, in the 1S-2S transition line \cite{fried1998bose,killian1998cold,killian20001}.
The efficacy and versatility of these methods are enhanced with transitions to higher-lying \textit{Rydberg} levels, since the loosely-bound Rydberg electron probes a region of space comparable to typical distances in a gas.
Its interaction with atoms within its orbit gives rise to spectroscopically detectable molecular lines for small densities which smear out into a broad shift as the particle number increases \cite{Greene2000,Fermi,Segre,Schlag,UltracoldChem}.
This approach has enabled measurements of pair correlations and of relative densities within a gas mixture \cite{whalenHeteronuclear2020,peperHeteronuclear2021a,dingCreation2020,whalenProbing2019,whalenFormation2019}.
At higher densities and larger principal quantum numbers, embedded Rydberg atoms in BECs or Fermi gases have served as ideal probes of polaron physics \cite{Demler,WhalenPoly}, measuring local density-density correlations and the effects of particle statistics \cite{camargoCreation2018a,sousRydberg2020}.
By increasing the principal quantum number it is possible to vary the spatial resolution from tens to a few hundreds of nanometers, a range inaccessible to typical optical wavelengths epitomizing the importance of Rydberg atoms in many-body settings.
Eventually, any further increase in the absorbed photon's energy ionizes the electron, raising the following question: can these \textit{near threshold} electrons also serve as probes of ultracold gases?

This question calls to mind the extended X-ray absorption fine structure spectroscopy (EXAFS) in solids, which uses highly energetic photoelectrons to probe the ionized atom's surroundings at nanometer resolution \cite{leeRMP1981,rehrRMP2000}.
This process yields photoionization cross-sections exhibiting a unique interference pattern, i.e. the photoionization fine structure, subject to the specific crystalline structure of the solid.
However, in ultracold atoms the typical length and energy scales, 100nm-$\mu$m and less than a few eV, are far different from those in solids, prohibiting the direct application of EXAFS theory.
In particular, the Coulomb force between the photoelectron and the ionized core can be completely ignored in EXAFS, but it plays a dominant role in the dynamics of slow photoelectrons.

\begin{figure}[t]
\includegraphics[width=1.\columnwidth]{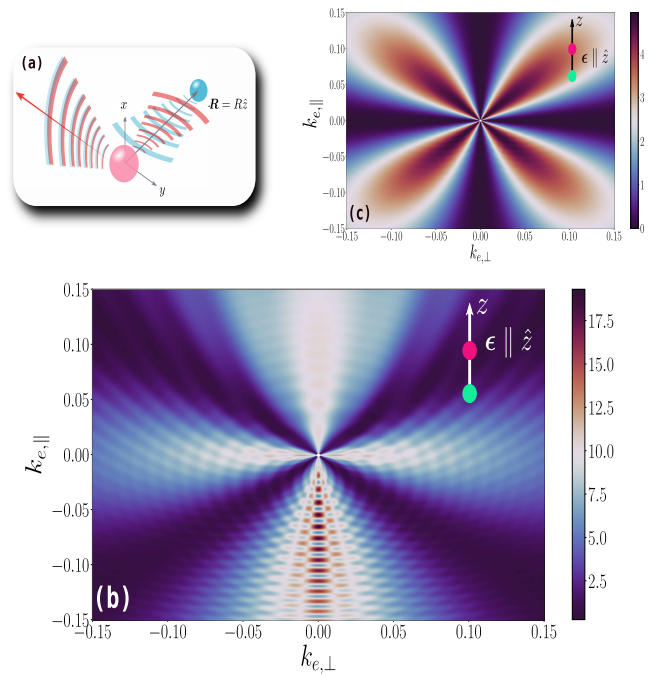}
\caption{(color online) (a) An illustration of the photoionization process, depicting the outgoing (red) and backscattered (blue) photoelectron. (b) and (c) contour plots of the differential photoionization cross-section of a Rb$(6p)$ atom in the presence of a perturber atom for linear (b) and circular (c) laser polarization.%The atoms are separated by $|{\boldsymbol R}|=400 ~a_0$ and $\hat R$ is parallel to the laser polarization axis $\boldsymbol{\epsilon}$.
}
\label{fig:fig1}
\end{figure}

In this letter, we develop a theoretical framework describing low-energy photoionization processes in the presence of ground state atoms.
Using this, we demonstrate that the coherent backscattering of the ionized photoelectron from a neighboring atom (see illustration in \cref{fig:fig1}(a)) imprints an oscillatory fine structure onto the photoionization cross-sections.
This fine structure exists in an energy interval slightly above threshold and contains information about the distance between atoms, the properties of the electron-atom and electron-ion subsystems, and the orientation of the atoms with respect to the polarization axis of the ionization laser field.
Therefore, this fine structure effect can be utilized as a structural probe of several aspects of an ultracold gas. 
Additionally, we show that, in the limit of highly kinetic photoelectrons, our formalism recovers the EXAFS theory, showcasing the general scope of our framework.
In order to focus on the general aspects of photoionization processes with neighboring atoms, we neglect spin and relativistic effects and we focus on the specific case of one nearby ground state atom of dominant importance.
To illustrate the potential utility of the low-energy photoionization fine structure, we propose a protocol to retrieve low-energy electron-atom phase shifts from measured photoionization fine structure spectra via a neural network method.
Our proposal utilizes a long-range Rydberg molecule as the initial state, permitting precise determination of the interatomic separation of the two relevant atoms.
Throughout this study we use atomic units, unless otherwise specified.

Our system consists of an unbound electron, an ionized atom, and a ground state ``perturber" atom  which is placed at a distance $\boldsymbol{R}=R\hat{z}$ relative to the ion.
The photoionization cross-sections of this three-body system are calculated by employing the generalized local frame transformation (GLFT) method, which treats the short-range potentials of the perturber and the ion on equal footing and has been successfully applied to describe long-range Rydberg molecules for electronic energies under the ionization threshold \cite{giannakeasprl2020,giannakeaspra2020}.
The effects of the short-range potentials are described by the electron-perturber scattering phase shifts $\delta_L$ and the atomic quantum defects $\mu_l$, where $L(l)$ denotes the orbital angular momentum of the photoelectron relative to the perturber (ionic core).
The GLFT theory elegantly expresses the dipole matrix elements relevant to the photoionization cross-section in terms of these parameters.
More specifically, the transition dipole matrix element between the initial and the $l- $th final state reads
\begin{equation}
	D^-_l(R)=\sum_{l'} e^{i(\eta_l+\pi\mu_{l'})}[\mathbb{1}-i \overline{K}(R)]^{-1}_{ll'}D_{l'}(R),
	\label{eq:eq1}
\end{equation}
where $\eta_l={\rm{ln}}2k_e/k_e-l\pi/2+{\rm{Arg}}\Gamma(l+1-i/k_e)$ is the Coulomb phase and $\overline{K}(R)$ is the {\it{smooth}} K-matrix describing the electron-perturber scattering process (for details see \cite{sm}).
Note that the final state in the definition of $D_l^-$ obeys incoming wave boundary conditions; the dipole matrix element $D_{l'}(R)$ corresponds to the case where the final state obeys standing wave boundary conditions, and is given by the relation
\begin{equation}
	D_{l'}(R)=\sum_{lLL'} [\delta_{l'l} +\pi \mathcal{B}_{l'm,LM}(R)[\mathcal{M}^{-1}]_{LL'} \mathcal{C}^T_{L'M,lm}(R)]d_{lm}.
	\label{eq:eq2}
\end{equation}
Here, $m(M)$ is the projection quantum number of $l(L)$.
The terms $\mathcal{B}_{l'm,LM}(R)$ and $\mathcal{C}_{L'M,lm}(R)$ control the admixture of $l$ and $L$ angular momenta by the perturber, while $M_{LL'}$ accounts for the electron-atom scattering (for details see \cite{sm}).
Finally, $d_{lm}=\braket{F_{lm}|\hat{d}|\psi_i}$ are the  dipole matrix elements describing transitions from the initial state $\psi_i(\boldsymbol{r})$ to an $l- $th final state $F_{lm}(\boldsymbol{r})=Y_{lm}(\hat{\boldsymbol{r}})[\cos \pi \mu_l f_l^c(\boldsymbol{r})-\sin \pi \mu_l g_l^c(\boldsymbol{r})]$, where $Y_{lm}(\hat{\boldsymbol{r}})$ are the spherical harmonics and $f_l^c(\boldsymbol{r}) ~[g_l^c(\boldsymbol{r})]$ denote the energy-normalized regular (irregular) Coulomb functions.
\cref{eq:eq2} describes the photoionization process that takes place near excited atom and it is valid only if the electron's orbit is smaller than the interatomic distance $\boldsymbol{R}$.
\cref{eq:eq1,eq:eq2} are the main general relations describing the motion of a photoelectron in the combined fields of an ionized and a neutral atom.

\cref{fig:fig1}(a) illustrates the coherent backscattering mechanism.
The red atom absorbs a photon and then releases a low-energy photoelectron which propagates outwards into the ionic core's Coulomb field with momentum $k_e$ (pink rays).
At distances $\boldsymbol{r}\approx\boldsymbol{R}$ part of the electron's wavefunction scatters off of the perturber (blue sphere) due to the electron-perturber interaction.
The backscattered wavefunction (blue ray) is phase-shifted by this collision, and accumulates an additional Coulomb phase as it returns to the source where it recombines with the part of the wavefunction that goes directly to the continuum.
This leads to an interference pattern that is imprinted in the cross-section as an oscillatory fine structure.

Such oscillations are clearly visible in the differential cross-section shown in \cref{fig:fig1} (b). 
For this calculation we consider specifically the photoionization of a Rb$(6p)$ atom in the neighborhood of a Rb$(5s)$ perturber \footnote{ We calculate the only input parameters -- the quantum defects $\mu_l$ and electron-perturber phase shifts -- using a non-relavatistic model potential \cite{Marinescu} and a two-electron R-matrix approach \cite{TaranaCurikLi,EilesHetero}.}.
Here, as well as in panel (c), we have fixed the interatomic distance $R=400~a_0$ and aligned the laser's polarization and interparticle axes.
The laser light is polarized linearly in panel (b) and circularly in panel (c). $k_{e,\parallel}$ ($k_{e,\perp}$) denotes the photoelectron momentum component parallel (perpendicular) to the interatomic axis.
For linearly polarized light the photoelectron is ejected to the continuum towards and away from the perturber, inducing enhanced interference fringes particularly on the negative axis.
For circular polarization, this strong interference pattern is heavily suppressed to the extent that the differential cross-section is smooth (see \cref{fig:fig1}(c))
In addition, the differential cross-section vanishes along the interatomic axis, which suggests that the ejected photoelectron does not collide with the perturber, yielding a cross-section virtually identical to that of a single atom.

\begin{figure}[t]
\includegraphics[width=0.98\columnwidth]{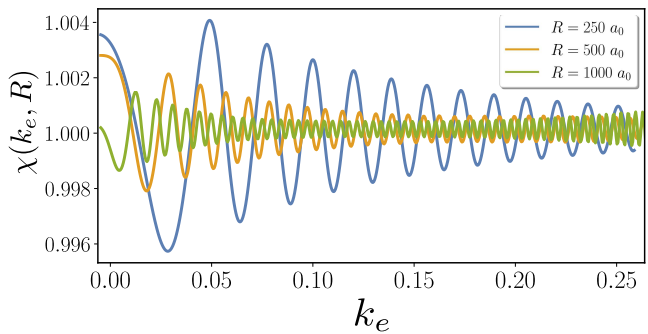}
\caption{(color online) Fine structure spectrum $\chi(k_e,R)$ of a $Rb(6p)$ atom near a perturber for $R=250~a_0~{\rm{(blue)}},~ 500~a_0~{\rm{(orange)}}$ and $1000~a_0~{\rm{(green)}}$ . The angle between the interatomic and the laser polarization axis is averaged over.}
\label{fig:fig2}
\end{figure}

After averaging over the solid angles, the oscillatory fine structure observed in \cref{fig:fig1}(b) is still observed in the total cross-section $\sigma(k_e,R)$, obtained for a specific interatomic separation $R$. 
The ratio of total cross-sections with and without the ground state atom defines the fine structure spectrum $\chi(k_e,R) = \sigma(k_e,R)/\sigma(k_e,\infty)$, isolates the physical impact of the backscattering mechanism.
Thus far we presented results for an idealized case where the polarization and interatomic axes are aligned in order to showcase the influence of neighboring atoms in the photoionization process.
However, in a dilute ultracold gas these two axes are not necessarily aligned and an average over all possible orientation must be considered.
We show that the fine structure $\chi(k_e,R)$ survives this averaging procedure in \cref{fig:fig2} manifesting the robustness of the photoionization fine structure effect.
Additionally, \cref{fig:fig2} shows the remaining dependence on $R$: the fine structure's amplitude (frequency) shrinks (grows) with increasing interatomic distances.

The  calculations in \cref{fig:fig2} bear a clear resemblance to x-ray photoionization fine structure, but a direct link between the GLFT theory and the established EXAFS formulas is stymied by the complicated angular momentum recoupling terms and Coulomb functions in \cref{eq:eq1,eq:eq2}.
Therefore, in the following we simplify these relations by considering a case where both atoms are initially in the ground state, have no quantum defects, and have only one dominant scattering channel.
For the latter we choose the $P$-wave channel since a low-energy shape resonance in the alkali atoms gives it a large contribution which can help to compare the relative effects of Coulomb and scattering phases in the photoionization process. 
We again align the linearly polarized laser with the interatomic axis.
Under these simplifications, the GLFT theory, combined with Jeffereys-Wentzel–Kramers–Brillouin (JWKB) approximation, permits us to analytically obtain an expression for the photoionization fine structure (for details see \cite{sm}),
\begin{equation}
    \chi(k_e,R)=1-9 \frac{K(R) \sin [\delta_P(\kappa)]}{\kappa^3R^2}\cos[2 \Phi(R) + \delta_P(\kappa)],
    \label{eq:eq3}
\end{equation}
where $\Phi(R)= \int^R dr K(r)$, $K(R)=\sqrt{k_e^2+2/R-9/4R^2}$ is the JWKB local momentum, $\delta_P(\kappa)$ denotes the electron-atom phase shift and $\kappa=\sqrt{2/R+k_e^2}$ indicates the momentum of the photoelectron relative to the perturber.
From \cref{eq:eq3} we observe that the peak-to-peak amplitude  of the fine structure falls off approximately as $\sim1/R$ for fixed $k_e$ and as $\sim 1/k_e^2$ for fixed R.

\begin{figure}[t]
	\includegraphics[scale=0.2]{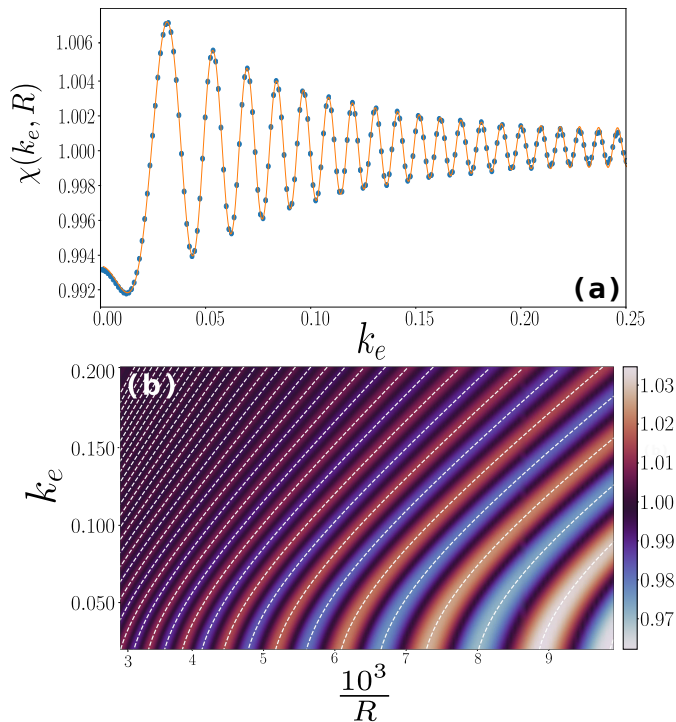}
\caption{(color online) (a) Comparison between the GLFT calculations (orange line) and the JWKB model (blue dots) for the $\chi(k_e,R)$ for $R=400~a_0$. Panel (b) shows the photoionization fine structure obtained via the GLFT theory. The white dashed lines correspond to the extrema of the photoionization fine structure (\cref{eq:eq3}). For both panels we photoionize from a Rb$(6s)$ state with a linearly polarized laser aligned with the interatomic axis and omit the quantum defects.}
\label{fig:fig3}
\end{figure}

\cref{fig:fig3} illustrates comparisons between \cref{eq:eq3} and the corresponding GLFT calculations for $R=400~a_0$.
Panel(a) shows the excellent agreement between the $\chi(k_e,R)$ obtained via the GLFT theory (orange line) and the predictions of \cref{eq:eq3} (blue dots).
\cref{fig:fig3}(b) depicts a contour plot of $\chi(k_e,R)$ calculated using the GLFT theory as a function of $R$ and $k_e$.
The white dashed lines trace out the extrema of $\chi(k_e,R)$ as predicted by \cref{eq:eq3}, which are again in excellent agreement with the numerical calculations over the entire parameter space.
Apart from the quantitative comparison, \cref{eq:eq3} transparently shows that the emergent interference pattern depends on the sum of the scattering phase shift $\delta_P(\kappa)$ and the photoelectron's phase accumulation in the Coulomb field, i.e. $\Phi(R)$.
% As intuitively seen in the physical picture of \cref{fig:fig1}(a), \cref{eq:eq3} suggests that the initial wave (red) accumulates a phase $\Phi(R)$ as it travels along the interatomic axis, before adding an additional phase, $\delta_P(\kappa)$, during the collision with the perturber.
% The latter produces the backscattered wave (blue), which travels towards the ionic core accumulating an additional $\Phi(R)$.
% Thus, the initial and backscattered waves are  $2\Phi(R)+\delta_P(\kappa)$ radians out of phase.
The theory of EXAFS yields a similar expression as \cref{eq:eq3}, although, since our results explicitly include the Coulomb field, are inherently far more general allowing the description of low- as well as high-energy photoelectrons.
The conceptual clarity of \cref{eq:eq3} suggests that measured photoionization fine structure in ultracold gases can be used to deduce the physical properties of the system and its constituent particles in an analogous way to EXAFS.

\begin{figure}[t]
\includegraphics[width=0.98\columnwidth]{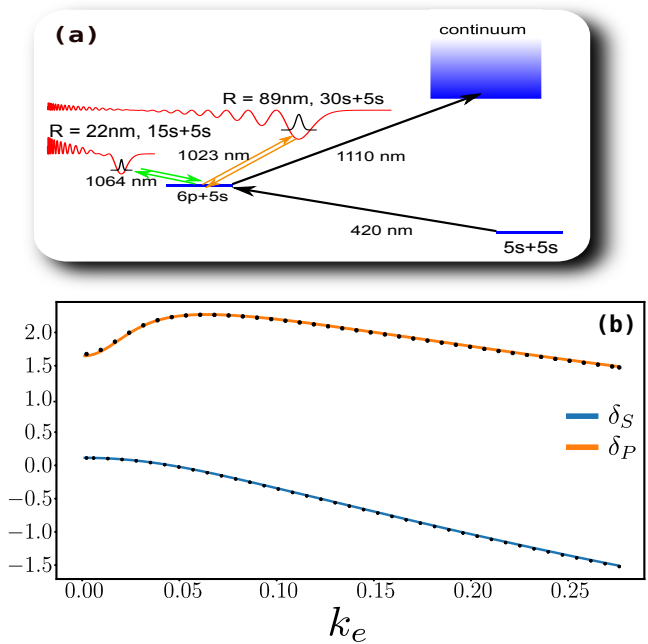}
\caption{(color online) a) The proposed experimental scheme to prepare a photoionization experiment involving a Rydberg molecule with fixed bond length.In the first step, two photons (black and green) are absorbed by an atom, resulting in an long-range Rydberg molecule with a well-defined bond length.The wavelength of the second laser selects the Rydberg level and resulting bond length.In a second step the Rydberg molecule is de-excited to a Rb$(6p)+$Rb$(5s)$ state, and subsequently the Rb$(6p)$ atom is photoionized. Such a scheme would allow the photoionization fine structure for fixed $R$ to be measured. (b) The S- (blue line) and P-wave (orange line) electron perturber phase shifts versus $k_e$. The black dots correspond to the predictions of the neural network.}
\label{fig:fig4}
\end{figure}

% The conceptual clarity of \cref{eq:eq3}, as well as the tremendous success of the analogous EXAFS process as a structural tool in solids, suggests a variety of potential metrology applications.
Indeed, as a proof of concept, we demonstrate here a scheme which could be used to extract low-energy scattering phase shifts from experimentally measured photoionization fine structure.
Our proposal is sketched in \cref{fig:fig4}(a).
A two-photon excitation scheme is employed to excite a long-range Rydberg molecule, whose molecular bond length depends on the Rydberg level and is used to define in advance of the photoionization step a fixed interatomic distance.
Here we consider a Rydberg molecule with bond length $R=1000~a_0$.
From this molecular state, a $V$-type photoionization protocol via the $\rm{Rb}(6p)+\rm{Rb}(5s)$ intermediate state is utilized.
This is a standardized technique applied in recent experiments \cite{engel2018observation}.
Additionally, Rydberg molecular states possessing a significant dipole moment could be used to align the molecule's internuclear axis with the polarization axis, as in a pendular state \cite{Rost}, although we do not rely on that here.

In order to extract the electron-atom phase shifts from a photoionization fine structure spectrum we use a neural network approach.
The neural network is trained on a large sample of artificial photoionization fine structure calculations where we have included all the atomic quantum defects with both $S-$ and $P-$wave electron-atom phase shifts.
A large ensemble $(\sim 10^3)$ of photoionization fine structure spectra is generated using "artificial" $S-$ and $P-$wave phase shifts which randomly deviate from the exact values by 20\%.
A fully connected feed-forward neural network is implemented to map $\chi(k_e,R)$ to the electron-atom phase shifts based on Julia's library Flux \cite{Julia-2017,Flux.jl-2018} (for details see \cite{sm}).
An excellent performance is reached after passing the complete data set $\sim 10^4$ times through the network.
As shown in \cref{fig:fig4}(b) the neural network predictions (black dots) are in excellent agreement with the exact $S-$ (blue line) and $P-$wave (orange line) electron-atom phase shifts, with the relative error being within $\approx 0.5 \%$.
Notice that the range of electron-atom momenta $\kappa=\sqrt{2/R+k_e^2}$ is bounded from below at $k_e=0$ due to finite $R$. For this reason the electron-atom phase shifts do not vanish at $k_e\to 0$.
We note that to date the most successful way to obtain these phase shifts has also involved long-range Rydberg molecules, but utilizing a fitting procedure where measured vibrational spectra of Rydberg molecules are compared with theoretically calculated spectra. The phase shifts employed in the theoretical calculation are then modified until the spectra agree \cite{deissObservation2020,PhasesFeyMeinert,DeSalvo2015,Camargo2016}.
Such a process has improved the values of key scattering parameters, such as the zero-energy scattering length and the position of the $P$-wave shape resonances, but considerable discrepancies and uncertainties still remain and it is not possible to reliably deduce the full energy-dependent scattering phase shifts.
The proposed photoionization scheme demonstrated here would avoid the ambiguities of previous approaches and also provide information about the phase shifts at arbitrary energy.

In summary, we have investigated photoionization in a dilute ultracold gas, where a nearby perturber atom causes interference fringes, that is, a fine structure in the atomic photoionization cross section.
The GLFT method permitted us to explicitly treat the Coulomb field and the electron-atom interactions within this three-body system. 
Apart from Rydberg molecules, this phenomenon can be observed in systems such as atoms trapped at individual lattice sites \cite{cardmannjp2021}. 
Since this fine structure is insensitive to the alignment of the polarization axis with the interatomic one, it can be observed also in weakly bound Feshbach molecules as well \cite{chinrmp2010}.
Finally, as a proof of concept we demonstrated that near threshold photoelectrons can be used to probe the electron-atom scattering interaction at low energies, using a Rydberg molecule to prepare a favorable initial state and a neural network to extract details about the electron-atom interaction from the fine structure of the photoionization spectra.
The full potential of this method will unfold in future work with the inclusion of spin-orbit coupling effects and multiple scattering off of several perturber atoms. 
Future extensions of our theory will permit to investigate the emerging diverse settings of Rydberg interacting systems, such as Rydberg-Rydberg \cite{sassmannshausenObservation2016,hollerithQuantum2019,hollerithMicroscopic2021,shafferUltracold2018a} and, very recently, Rydberg-ion, bound states \cite{deissLongRange2021,duspayevLongrange2021a,zuberSpatial2021}, using the fine structure of photoionization and its variants as a sensitive probe.

\begin{acknowledgments}
The authors are thankful to F.Meinert for helpful discussions.
FR acknowledges the financial support by the U.S. Department of Energy (DOE), Office of Science, Basic Energy Sciences (BES) under Award No. DE-SC0012193.
LA acknowledges the financial support from the European Union’s Horizon2020 research and innovation project DeepCube, under grant agreement number 101004188.
The numerical calculations have been performed using NSF XSEDE Resource Allocation No. TG-PHY150003.
\end{acknowledgments}

\bibliography{Thesis_bib.bib}

%%%%%%%%%% Merge with supplemental materials %%%%%%%%%%
\clearpage
% \pagebreak
% \begin{widetext}
\onecolumngrid
\begin{center}
\textbf{\large Supplemental Material: Probing ultracold gases using photoionization  fine structure}
\end{center}
\twocolumngrid
% \end{widetext}
%%%%%%%%%% Merge with supplemental materials %%%%%%%%%%
%%%%%%%%%% Prefix a "S" to all equations, figures, tables and reset the counter %%%%%%%%%%
\setcounter{equation}{0}
\setcounter{figure}{0}
\setcounter{table}{0}
\setcounter{page}{1}
\makeatletter
\renewcommand{\theequation}{S\arabic{equation}}
\renewcommand{\thefigure}{S\arabic{figure}}
\renewcommand{\bibnumfmt}[1]{[S#1]}
\renewcommand{\citenumfont}[1]{S#1}
%%%%%%%%%% Prefix a "S" to all equations, figures, tables and reset the counter %%%%%%%%%%

% \section{Supplemental material}
In this supplementary material additional information is provided on various aspects of the main manuscript.
In summary, these extra details cover the following points:
\begin{itemize}
    \item Generalized local frame transformation theory for photoionization: key formulas for the derivation of the dipole matrix elements
    \item Derivations and approximations employed for Eq.(3) in the main text
    \item Details on the neural network fitting procedure used in Fig.4(c) in the main text
\end{itemize}

\section{Generalized local frame transformation for photoionization processes in ultracold gases}
In order to describe the photoionization process discussed in the main manuscript the generalized local frame transformation (GLFT) theory is employed.
The additional details given below are sufficient in order to compute the photoelectron's dipole matrix elements (see Eq.(1) and (2) in the main text).
Following the prescription given in Ref.\cite{giannakeaspra2020} we derive the relations of $\bar{K}_{l'l}(R)$ (see \cref{eq:s5}), $\mathcal{C}^T_{L'M,lm}(R)$ (see \cref{eq:s6a}) and $\mathcal{B}^T_{L'M,lm}(R)$ (see \cref{eq:s6b}) used in Eq.(1) and (2) of the main text.

\subsection{Quantum defect shifted Coulomb Lippmann-Schwinger equation and the smooth K-matrix}
%Based on the experimental protocol of Fig.1(a), after the creation of $^{85}Rb_2$ Rydberg molecule with bond length $R$, follows a de-excitation to $Rb(6p)+Rb(5s)$ state where the excited atom $Rb(6p)$ is immediately photoionized.
We study the photoionization of an atom $A$ in the presence of a bystander atom $B$.
The photoelectron is ejected into the continuum with energy $\epsilon=k_e^2/2$ experiencing the Coulomb attraction and the residue interaction of the ionic core, i.e. $A^+$.
At distances $R$ with respect to $A^+$, the photoelectron collides with a ground state atom $B$ via the electron-atom polarization potential.
As was shown in \cite{giannakeaspra2020}, the wave function of the electron, which is centered at the ionic core, can be expressed as a linear combination of the regular and irregular quantum-defect-shifted (QDS) Coulomb functions at positive energies.
Note that throughout the supplemental material we use atomic units unless otherwise specified.
\begin{equation}
	\label{eq:s1}
	\Psi_{lm}(\bm r, R)=\sum_{l'} Y_{l'm}(\hat r) [F_{l'}( r) \tilde \delta_{l'l}-G_{l'}( r)	\overline{K}_{l'l}(R)],
\end{equation}
 where $\tilde{\delta}_{l'l}$ is Kronecker's delta, $Y_{lm}(\hat r)$ correspond to the spherical harmonics centered at $A^+$ with $l$ and $m$ being the orbital and azimuthal angular momentum relative to the ionic core, respectively.
 Note that \cref{eq:s1} is a valid approximation since the coupling of electron's angular momenta and spins with the ionic core's corresponding degrees of freedom are neglected.
 The term $\overline{K}_{l'l}(R)$ refers to the smooth K-matrix and $F_{l'}(r)$ $(G_{l'}( r))$ is the regular (irregular) QDS Coulomb function.
The energy-normalized QDS pair of solutions are related to the conventional regular and irregular Coulomb functions ($f_l^c,~g_l^c$) according to
\begin{subequations}
	\begin{align}
		\label{eq:s2}
		F_l(r) =  f_l^c(r)\cos\pi\mu_l -  g_l^c(r)\sin\pi\mu_l,\\
		G_l(r) =   f_l^c(r)\sin\pi\mu_l + g_l^c(r)\cos\pi\mu_l, \label{eq:s2}
	\end{align}
\end{subequations}
where $\mu_l$ denotes the $l-$th atomic quantum defect and it parametrizes the impact of the residue potential of $A^+$ on the electron's motion.
The QDS pair of solutions $(F_l,~G_l)$ are related to the $\overline{K}$-matrix, i.e. the smooth K-matrix, their Wronskian evaluated at distance $r$ is given by the expression $\mathcal{W}\{F_l,G_l\}_r=2/(\pi r^2)$.
In addition, \cref{eq:s1} obeys the following Lippmann-Schwinger equation:
\begin{equation}
	\label{eq:s3}
	\ket{\Psi_{lm}(R)} = \ket{{F}_{l}(R)} + \hat {G}_C^{\rm{QDS}}\hat V_B\ket{\Psi_{lm}(R)},
\end{equation}
where $\hat{G}_C^{\rm{QDS}}$ is the quantum-defect-shifted {\it principal-value} Coulomb Green's function and $V_B$ is the electron-perturber interaction potential.
The QDS Green's function reads 
\begin{equation}
	\label{eq:s4}
	G_C^\text{QDS}(\bm r,\bm r') = \pi\sum_{lm}Y_{lm}^*(\hat r)F_l(r_<)G_l(r_>)Y_{lm}(\hat r'),
\end{equation}
where $r_< = \text{min}(r,r')$ and $r_> = \text{max}(r,r')$.

As in Ref.\cite{giannakeaspra2020}, the $\overline{K}$-matrix is given by 
\begin{equation}
		\overline{K}_{l'l}(R) = -\pi\sum_{LL'}\mathcal{C}_{l'm,LM}(R)[\mathcal{M}^{-1}]_{LL'}\mathcal{C}_{L'M,lm}^T(R).\label{eq:s5}
\end{equation}
where $L$ and $M$ denote the orbital and azimuthal angular momentum with respect to the ground state atom B.
The matrix elements $\mathcal{C}_{l'm,LM}(R)$ and $\mathcal{M}_{L'L}$ can be expressed in terms of the local frame transformations $\mathcal{U}^T_{lm,LM}(R,\mu_l)$,$\mathcal{V}^T_{lm,LM}(R,\mu_l)$, $\mathcal{J}^T_{lm,LM}(R)$ and $\mathcal{N}^T_{lm,LM}(R)$ according to the following relations:
\begin{equation}
	\mathcal{C}^T_{LM,lm}(R)=-\frac{\tan \delta_{L}}{\pi}\mathcal{U}^T_{LM,lm}(R,\mu_l),
	\label{eq:s6a}
\end{equation}
where $\tan \delta_L$ represents the electron-atom scattering phase shift.

\begin{widetext}
	\begin{align}
		&\mathcal{M}_{L'L} = -\frac{1}{\pi}\tan\delta_{L'} \tilde \delta_{L'L}+\frac{\tan\delta_{L'}\tan\delta_{L}}{\pi}\sum_{lm}\big[\mathcal{J}^T_{L'M,lm}(R)\mathcal{N}_{lm,LM}(R) - \mathcal{U}^T_{L'M,lm}(R,\mu_l)\mathcal{V}_{lm,LM}(R,\mu_l)\big].\label{eq:s7}
	\end{align}
\end{widetext}

Also, in Eq.(2) of the main text the dipole matrix elements for standing wave boundary conditions depend on the $\mathcal{B}_{lm,LM}(R)$ matrix elements which read
\begin{equation}
	\mathcal{B}_{lm,LM}(R)=-\frac{\tan \delta_{L}}{\pi}\mathcal{V}_{lm,LM}(R,\mu_l).
	\label{eq:s6b}
\end{equation}
\subsection{Local frame transformations}
The relations of the local frame transformations are:
\begin{subequations}
	\begin{align}
		\begin{Bmatrix} \mathcal{J}_{LM,lm}^T(R) \\ \mathcal{N}_{LM,lm}^T(R) \end{Bmatrix}
		&=
		\sqrt{\frac{2L+1}{2l+1}}\sum\limits_{\beta=0}^\infty i^{L + \beta -l}(2\beta + 1)\times\nonumber \\
		&\times C_{L0\beta0}^{l0}C_{LM\beta 0}^{lm}
		\begin{Bmatrix}
			j_\beta(\kappa R)\\
			n_\beta(\kappa R)\\
		\end{Bmatrix},\label{eq:s8}
	\end{align}
\end{subequations}
where $\kappa$ is the momentum of the electron relative to the ground state atom B and it obeys the relation $\kappa=\sqrt{2/R+k_e^2}$.
The terms $C_{j_1m_1,j_2m_2}^{j_3m_3}$ are the Clebsch-Gordan coefficients relating angular momenta pairs $(L,M)$ and $(l,m)$.
The pair of functions ($j_\beta, n_\beta$) indicate the spherical Bessel and Neumann functions.
Also, for $\mathcal{U}^T_{LM,lm}(R,\mu_l)$ and $\mathcal{V}^T_{LM,lm}(R,\mu_l)$ which are used in \cref{eq:s6a,eq:s7,eq:s6b} we have the relations
\begin{equation}
	\begin{aligned}
		\mathcal{U}^T_{LM,lm}(R,\mu_l)=\frac{\pi R^2}{2}&\bigg[\mathcal{J}_{LM,lm}^T(R) \mathcal{W}\{F_{lm},g^0_{lm}\}_R\\
		&-\mathcal{N}_{LM,lm}^T(R) \mathcal{W}\{F_{lm},f^0_{lm}\}_R\bigg] \label{eq:s9}
	\end{aligned}
\end{equation}

\begin{equation}
	\begin{aligned}
		\mathcal{V}^T_{LM,lm}(R,\mu_l)=\frac{\pi R^2}{2}&\bigg[\mathcal{J}_{LM,lm}^T(R) \mathcal{W}\{G_{lm},g^0_{lm}\}_R\\
		&-\mathcal{N}_{LM,lm}^T(R) \mathcal{W}\{G_{lm},f^0_{lm}\}_R\bigg],\label{eq:s10} 
	\end{aligned}
\end{equation}
where $\mathcal{W}\{\cdot,\cdot\}_R$ refers to the Wronksian evaluated at distance $R$.The pair of functions $[f^0_{lm}(\boldsymbol{r}),g^0_{lm}(\boldsymbol{r})]$ indicate the energy-normalized free-space solutions at momenta $\kappa=\sqrt{2/R+k_e^2}$ and they are proportional to the spherical Bessel and Neumann functions, respectively.
Note that the frame transformation in \cref{eq:s10} is different from the one used in Ref.\cite{giannakeaspra2020}.
Namely, in \cref{eq:s10} we employ the irregular QDS Coulomb functions since we assume that the electron possess positive energies.

\section{Simplified GLFT theory within Jeffereys-Wentzel–Kramers–Brillouin approximation}
As discussed in the main text, we derive Eq.(3) under a simplified set of conditions where a Rb$(6s)$ atom is photoionized by a linearly polarized light.
We neglect for simplicity the atomic quantum defects and the S-wave electron-atom phase shift.
In addition, we assume that the polarization and interatomic axis are aligned in the $z-$direction with the ground state atom Rb$(5s)$ placed at a distance $R$.
Under these considerations, we substitute \cref{eq:s5,eq:s6a,eq:s7,eq:s6b} into Eq.(3) and for final states with $(l_f,m_f)=(1,0)$ the corresponding dipole matrix elements $D_l(R)$ read:

\begin{equation}
\begin{aligned}
   	D_l(R)&=[\tilde{\delta}_{l,l_f} + Y^*_{l_fm_f}(\hat{R}) \partial_R g^c_{l_f}(R)A \\
   	&\times\partial_R f^c_{l}(R)Y_{lm}(\hat{R}) ]d_{lm}, ~~\rm with  \\ 
    A&=\pi \tan \delta_P(\kappa) [-\frac{\kappa^3}{6\pi}+\tan \delta_P(\kappa) \Delta(R)]^{-1},  \label{eq:s12}
\end{aligned}
\end{equation}
note that the preceding relation refers to the dipole matrix elements whose final states obey standing wave boundary conditions. $\kappa$ is the momentum of the electron relative to the ground state atom and is given by the expression $\kappa=\sqrt{2/R+k_e^2}$.$d_{lm}$ refer to the dipole matrix elements of a single atom where the final state obeys standing wave boundary conditions.
The term $\Delta(R)$ in the preceding equations reads:
\begin{equation}
    \Delta(R)=\sum_{lm}\mathcal{J}^T_{10,lm}(R)\mathcal{N}_{lm,10}(R) -\mathcal{U}^T_{10,lm}(R,\mu_l)\mathcal{V}_{lm,10}(R,\mu_l),
    \label{eq:s13}
\end{equation}

Using the \cref{eq:s12}, the dipole matrix elements which obey incoming boundary conditions read:
\begin{equation}
\begin{aligned}
    D^-_l(R)=&e^{i \eta_l}\big[\tilde \delta_{l,lf} +\frac{A}{1+i A\beta}Y_{l_f,m_f}^*(\hat{R})\partial_R[g^c_{l_f}(R) \\
    &\ldots-if^c_{l_f}(R)]\partial_Rf^c_l(R)\big], ~~\rm{with}~ \\
    &\beta = \sum_{lm} |Y_{lm}(\hat{R}) \partial_R f^c_l(R)|^2.\label{eq:s14}
    \end{aligned}
\end{equation}

Numerical calculations suggest that $\Delta(R)$ and $\beta$ exhibit fast oscillations with respect to the photoelectron's energy around zero and $\frac{\kappa^3}{6\pi^2}$, respectively.
Therefore,  \cref{eq:s14} can be further simplified by considering the following two approximations: (i) $\Delta(R)\approx0 $ and (ii) $\beta\approx\frac{\kappa^3}{6\pi^2}$.
In addition, the regular and irregular Coulomb functions, i.e. $f^c_l(R)$ and $g_l^c(R)$, can be evaluated in the Jeffereys-Wentzel–Kramers–Brillouin (JWKB) approximation.
Utilizing these approximations, the photoionization fine structure, i.e. the ratio of the total cross-section with and without the ground state atom, has the following form:
\begin{equation}
\chi(k_e,R)=1 +9 \frac{\Gamma^2}{\kappa^3R^2 K(R)}\sin \delta_P(\kappa) \sin [2\Phi(R)-2t + \delta_P(\kappa)],
    \label{eq:s15}
\end{equation}
where $\Gamma^2=K^2(R)+[1/R+\partial_RK(R)/K(R)]^2$ and $t=\arctan[K^2(R)/(\Gamma^2-K(R)^2)]$.$K(R)$ denotes the JWKB local momentum which contains only the Coulomb potential, $\Phi(R)$ refers to JWKB phase accumulation in the presence of the Coulomb field, and $\delta_P(\kappa)$ is the P-wave electron-atom phase shift.
Recall that \cref{eq:s15} depends only on P-wave phase shift since we neglected the S-wave scattering process.
From numerical tests we observe that at large $R$ the phase $t\approx\pi/2$ and $\Gamma\approx K(R)$.These approximations, although not necessary, simplify further \cref{eq:s15} yielding the expression used in the main text.
\begin{equation}
    \chi(k_e,R)= 1-9\frac{K(R)}{\kappa^3 R^2} \sin \delta_P(\kappa) \sin[2\Phi(r)+\delta_P(\kappa)].
\end{equation}
\section{Details of the Neural Network fitting procedure}
We generated $n=10^3$ samples by adding random variation on the order of $20\%$ to the exact values of the electron-atom phase shifts, and using these to compute total cross-sections.
Subsequently, the samples were split between training ($n_{\rm train} = 0.7n$), validation  ($n_{\rm val} = 0.1n$) and test ($n_{\rm test} = 0.2n$) data.
A fully connected feed-forward neural network is implemented to map $\chi(k_e,R)$ to the electron-atom phase shifts simultaneously based on Julia's library \cite{Julia-2017} Flux \cite{Flux.jl-2018}.
The mean square error is used as the training loss and ReLU \cite{pmlr-v15-glorot11a}  as the activation function between hidden layers.
The initial parameters of the network are generated using the XAVIER initialization \cite{pmlr-v9-glorot10a} and the optimization of the network (update of parameters) is done using the ADAM algorithm \cite{kingma2017adam}.
Also, in order to reduce the computational cost we applied mini-batch optimization \cite{KDD-2014-LiZCS} with a batch size of 128.
An excellent performance is reached after passing through the network the complete data set $\sim 10^4$ times.
Note that the final optimized network applied to the test data is the one with the lowest validation loss.
% \footnote{http://de.wikipedia.org/wiki/Welt}{Code and data are available publicly}.

%\bibliography{Thesis_bib.bib}
\end{document}